\documentclass[preprint,twocolumn]{article}
\usepackage{amsmath}
\usepackage{amsfonts}
\usepackage{graphicx}
\usepackage{xcolor}
\usepackage{multicol,caption}
\usepackage{caption}
\usepackage{sectsty}
\sectionfont{\large}

\usepackage[hyphens]{url}
\usepackage{hyperref}
\usepackage[hyphenbreaks]{breakurl}

\hypersetup{breaklinks=true}

\usepackage{cite}

\begin{document}

\title{Magnon bands and transverse transport in a proposed two-dimensional $Cu_2F_5$ ferrimagnet}
\author{Pedro G. de Oliveira$^a$ and Antônio S. T. Pires$^a$}
\date{}

\twocolumn[
  \begin{@twocolumnfalse}
    \maketitle
    \begin{center}
	\textit{$^a$Department of Physics, Federal University of Minas Gerais, Belo Horizonte, MG, Brazil.} 
	\end{center}
	\vspace{\baselineskip}

    \begin{abstract}
      The copper fluoride $Cu_2F_5$ is a proposed stable compound that can be seen as a layered lattice of $S=1$ and $S=1/2$ sites, corresponding to copper ions. Intending to cast light on the transport properties of ferrimagnetic magnons, we use the linear spin wave approach to study the magnon band structure of the 2D lattice in a ferrimagnetic off-plane order, as well as the transverse transport of magnons in the crystal bulk. A magnetic field or temperature gradient can induce transverse (Hall-like) transport within the linear response theory generated by the Berry curvature of the eigenstates. As in most cases involving magnons, the Berry curvature is related to Dzyaloshinskii-Moriya interactions between next-near-neighbors. The band structure of the system is non-degenerate and the transport coefficients are non-null. The novel and interesting feature of the system is that, within certain conditions, the transport coefficients versus temperature curves are non-monotonic and present a sign change, opening the exciting possibility of controlling the transverse magnon flow direction and magnitude with the temperature change.
    \end{abstract}
    
\vspace{\baselineskip}
\vspace{\baselineskip}

  \end{@twocolumnfalse}
]

\small

\section{Introduction}

When a crystal has magnetic order, perturbations in this order propagate through the crystal in the form of \textit{spin waves}. From a quasiparticle point of view, these spin waves are called \textit{magnons}, which are chargeless bosons with definite spin, momentum and energy. In the past years, there has been a great effort to study the creation and manipulation of magnons \cite{Bracher2017,Wu2018,Bai2015}, as well as their interaction with other particles or quasiparticles \cite{Kittel1958,Cao2015_polaritons,Zhang2014,Bhoi2019,dip2}, including the formation of hybrid modes \cite{Takahashi2016,Agrawal2013,Rockriegel2014,Delugas2023,Ghirri2023_preprint,Mook2023,
Ghosh2023, Tabuchi2014}. These efforts rely on the exciting possibility of using magnons' spin degree of freedom to transport information in a field known as \textit{magnonics} (magnon spintronics) \cite{Kruglyak2010,Chumak2015,Wang2021}. Magnonics has a significant advantage over electron-based spintronics: magnons are naturally chargeless, do not present Joule heating, and can propagate large distances with low dissipation.

Within this context, it became essential to study the transport properties of magnons in different lattices and magnetic orders. It is theoretically established that a magnetic field or temperature gradient can generate longitudinal and transverse (Hall-like) transport of magnons in the bulk of a crystal \cite{Fujimoto2009,Katsura2010,Matsumoto2014,Han2017,Nakata2017FM}. These transport effects have been intensely studied in ferromagnets (FM) \cite{Katsura2010,Matsumoto2014,Han2017,Nakata2017FM,Matsumoto2011,Matsumoto2011_2,Zhang2013,Mook2014,Mook2016,Mook2016_2,
Cao2015,Chisnell2015,Zarzuela2016,kovalev2016,Owerre1,Owerre3,Owerre2,
Pantaleon2018,Le2019,Pires2021_2,deOliveira2023_Lieb} and antiferromagnets (AFM) \cite{Chen2014,Kubler2014,Rezende2016,Cheng2016,Zyuzin2016,Li2016,Nakata2017,
Owerre6,Owerre8,Laurell2018,Doki2018,Zhang2018,Zelezny2018,Mook2019,
Lu2019,Kim2019,Kawano2019,Li2020,Smejkal2020,Azizi2020,Pires2020,Pires2021,Bonbien2021,Zhang2022,Kondo2022,
Liu2023,deOliveira2023_2}. The theoretical study of these thermomagnetic properties is well developed, and nowadays, we can describe magnonic analogs for the \textit{quantum Hall effect} and \textit{quantum spin Hall effect} of electrons \cite{Nakata2017FM,Nakata2017}, among other exotic transport effects. Experimental evidence has demonstrated the existence of transverse transport in the bulk of 3D and 2D lattices \cite{Onose2010,Ideue2012,Hirschberger2015,HirschbergerScience2015,Shiomi2017}. 

It has also been established that magnonic systems can show protected edge or surface states, which are robust against non-magnetic impurities. That comes from the well-known bulk-edge correspondence for topological systems, which indicates the existence of \textit{topological magnon insulators} (TMI) \cite{Wang2021,Zhang2013,Zarzuela2016,Owerre1,McClarty2022}. That term has been used as an analogy to electronic topological insulators. Like electronic topological insulators, the TMI are characterized by topological indices like the Chern number or $\mathbb{Z}_2$ invariant \cite{Nakata2017FM,Nakata2017,Kondo2019,Kondo2020}. Despite that, the bosonic nature of magnons excludes the existence of a Fermi level, so the system is not an insulator \textit{strictu sensu}. Topological and Hall-like effects of magnons are related to the spin-orbit coupling of the crystal lattice, and both rely on the Berry curvature of the system, which acts like a fictitious magnetic field and imparts helical movement to the magnons.

Although widely studied in FM and AFM systems, the Hall transport of magnons has not been well investigated in ferrimagnetic (FiM) lattices \cite{Park2020,Nakata2021}. With that in mind, in this work we investigate the magnon Hall transport in the 2D layers of the proposed copper fluoride complex $Cu_2F_5$, whose stability was predicted recently with first-principle methods \cite{Rybin2021,Korotin2021,Korotin2023}. In this crystal, copper ions have two different spin states ($S=1$ and $s=1/2$), forming a ferrimagnetic system that we investigate using the spin wave approach. The spin wave formalism for FiM and AFM lattices is identical, and the reader is referred to Ref. \cite{deOliveira2023_2} for a detailed description of this formalism.

This paper is organized as follows. We present the crystal structure and the Hamiltonian, which models the 2D magnetic lattice, in Section \ref{section_model}. In Section \ref{section_berry} we discuss the Berry curvature of magnon bands and its implications for the transverse transport of magnons. In Section \ref{section_results} we present the results of the systems's band structure and transverse transport coefficients. In Section \ref{section_conclusion} we make our final remarks.

\section{Model}
\label{section_model}

\begin{figure}[h!]
\centering
\includegraphics[width=0.45\textwidth]{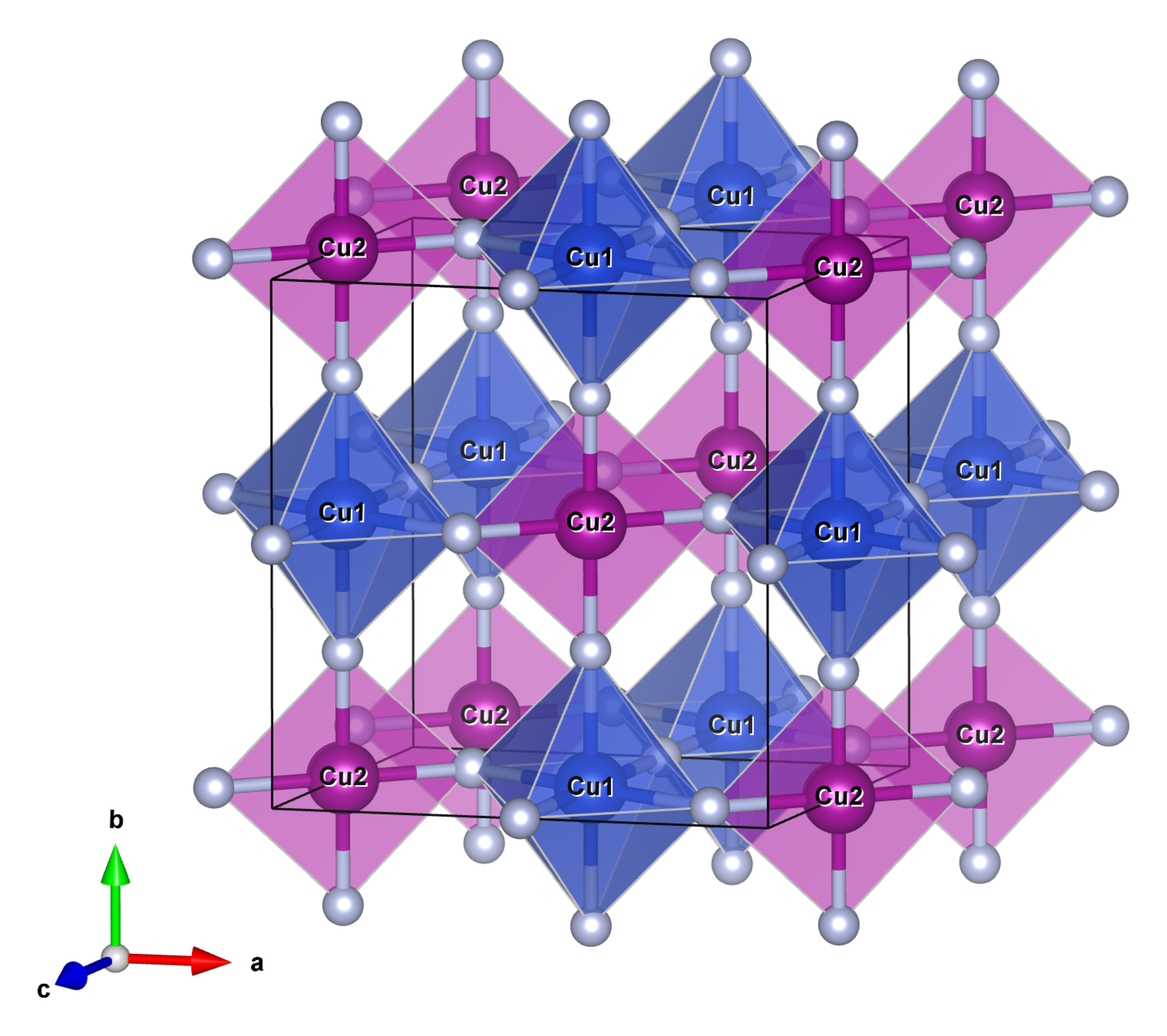} \caption{The crystal structure of the proposed $Cu_2F_5$ lattice. $Cu$ ions inside the blue octahedra ($Cu1$) have spin $S=1$. $Cu$ ions inside the magenta plaquettes ($Cu2$) have spin $s=1/2$. Reproduced with permission
from Ref. \cite{Korotin2021}.} \label{figure_crystal}
\end{figure}

The proposed $Cu_2F_5$ crystal is composed of $CuF_6$ distorted octahedra and $CuF_4$ plaquettes, as shown in Figure \ref{figure_crystal} \cite{Rybin2021,Korotin2021}. The inequivalent $Cu$ ions form a magnetic crystal. We call $Cu1$ the $S=1$ ions in the center of the octahedra, and $Cu2$ the $s=1/2$ ions in the center of the plaquette. The most energetically favorable configuration is a G-type ferrimagnet, and DFT+U calculations show that the 3D crystal can be seen as a layered structure with the interlayer exchange parameter five times smaller than the intralayer ones \cite{Korotin2021}. That inspires us to study the 2D layers from a spin wave point of view. We chose a ferrimagnetic off-plane spin configuration, with the $S=1$ spins pointing in the $+\mathbf{\hat{z}}$ direction and the $s=1/2$ in the $-\mathbf{\hat{z}}$ direction. The stacked layers form a ferrimagnetic C-type configuration, and the magnetic lattice has two inequivalent sites. That is not the most stable configuration, but it is much simpler than the aforementioned G-type FiM, which has four inequivalent sites.

The exchange model which stabilizes the the layer's spin order comprises a FM exchange bond between $Cu1$ sites and an AFM exchange bond between $Cu1-Cu2$ sites. We add a Dzyaloshinskii-Moriya interaction (DMI) between next-near-neighbors (NNN) sites and a single-ion anisotropy (SIA) in the $z$-direction. The DMI is responsible for the transverse transport, and the SIA stabilizes the off-plane configuration. The Hamiltonian of the model is:

\begin{align}
H=&-J_{1}\sum_{\left\langle ij\right\rangle }S_{i}\cdot S_{j}+J_{2}\sum_{\left\langle ij\right\rangle }S_{i}\cdot s_{j} \nonumber \\
  &+D\sum_{\left\langle \left\langle ij\right\rangle \right\rangle }\nu_{ij}\left(S_{i}^{x}s_{j}^{y}-S_{i}^{y}s_{j}^{x}\right) \nonumber \\
  &-A\sum_{i}\left[\left(S_{i}^{z}\right)^{2}+\left(s_{i}^{z}\right)^{2}\right] \label{hamiltonian_main}
\end{align}

\begin{figure}[h!]
\centering
\includegraphics[width=0.5\textwidth]{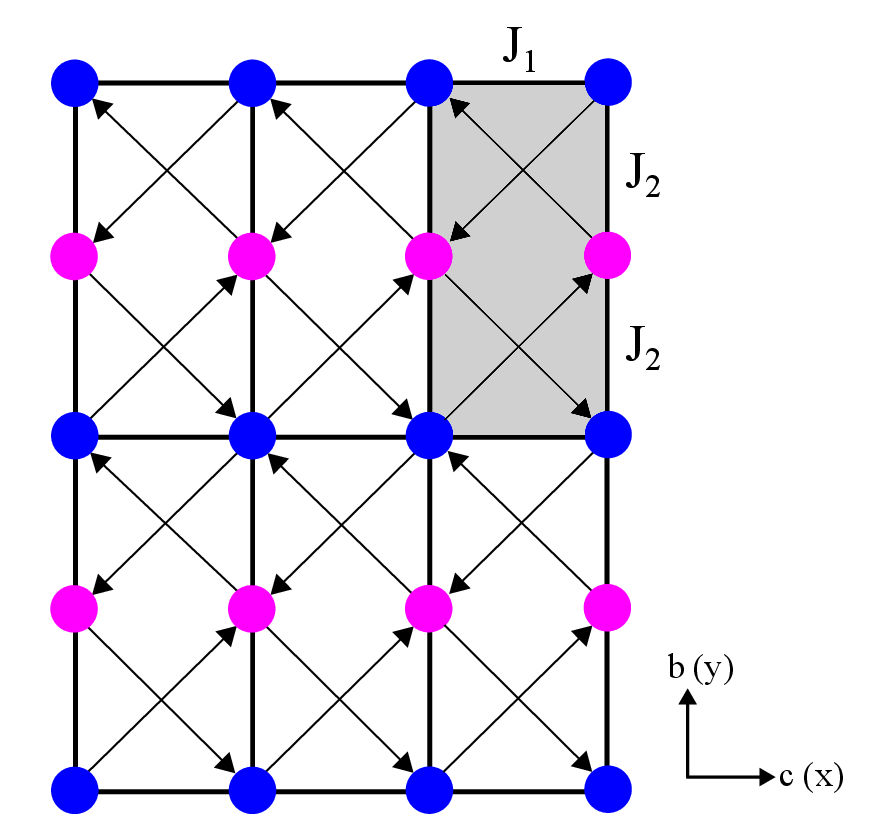} \caption{2D model studied in this paper, corresponding to a layer in the b-c plane of the crystal structure in Figure \ref{figure_crystal}. Exchange interactions are represented as $J_1$ and $J_2$. A Dzyaloshinskii-Moriya interaction exists between NNN with $\nu_{ij} = +1 (-1)$ along (against) the arrow. Blue sites have spin $S=1$, and magenta sites, $s=1/2$. The gray region corresponds to the unit cell.} \label{figure_lattice}
\end{figure}

The upper (lower) case $S_i$ ($s_i$) operator denotes the spin operators for $S=1$ ($s=1/2$) sites. The first term ($J_1>0$) represents the FM exchange between $S=1$ ($Cu1$) sites, and the second term ($J_2>0$) represents the AFM exchange between $S=1$ and $s=1/2$ ($Cu1-Cu_2$) sites. Both interactions happen between near-neighbors (NN). The third term is the DMI between NNN sites, where $\nu_{ij}= \pm 1$, following the arrow convention in Figure \ref{figure_lattice}. The last term is the SIA. We note that the SIA between 1/2 spins is ineffective \cite{deOliveira2023_2}, so only the first term inside the square brackets needs to be considered.

We use the linearized Holstein-Primakoff representation for up/down spins:

\begin{align}
S_{i}^{+}  &  =\sqrt{2S}\,a_{i}\,\,,\,\,\,S_{i}^{-}=\sqrt{2S}\,a_{i}^{\dagger
}\,\,,\,\,S_{i}^{z}=S-a_{i}^{\dagger}a_{i}\nonumber\\
s_{i}^{+}  &  =\sqrt{2s}\,b_{j}^{\dagger}\,\,,\,\,\,\,\,s_{i}^{-}=\sqrt{2s}\,%
b_{j}\,\,,\,\,\,\,\,s_{i}^{z}=-s+b_{j}^{\dagger}b_{j}
\label{HP}%
\end{align}

\begin{figure*}[t!]
\centering
\includegraphics[width=0.9\textwidth]{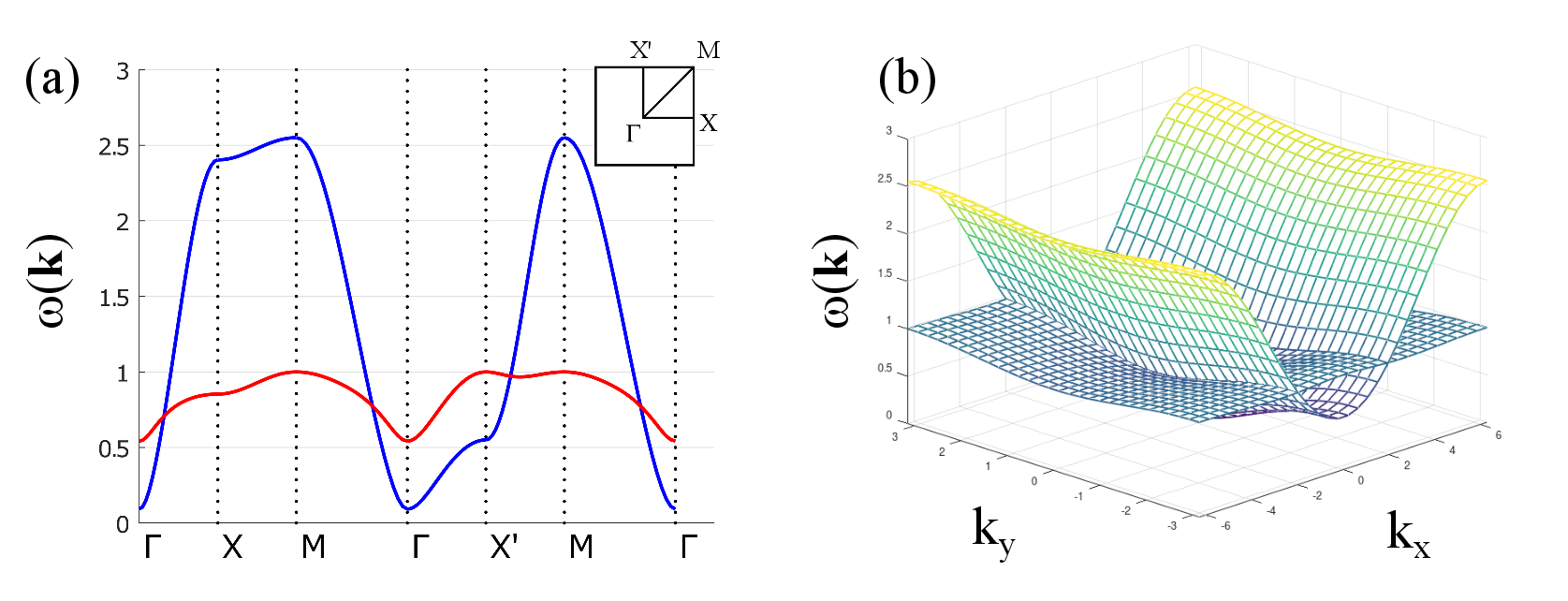} \caption{(a) Band structure of the system (between high symmetry points) when $J_{1}=J_{2}=1.0$, $A=0.1$, and $D=0.2$. The blue (red) line corresponds to the $\omega_{\downarrow (\uparrow)} \left( \mathbf{k}  \right)$ band. (b) Band structure throughout the whole Brillouin zone.} \label{figure_bands}
\end{figure*}

So, the Hamiltonian is written in terms of the magnon creation and annihilation operators in configuration space. This representation can be applied to collinear AFM and FiM with off-plane spins. Performing a Fourier transform enables us to write the momentum-space 4x4 Hamiltonian in the block matrix form:

\begin{equation}
H_{k}=\psi_{k}^{\dagger}\left(\begin{array}{cc}
H_{k}^{I} & 0\\
0 & H_{k}^{II}
\end{array}\right)\psi_{k}
\end{equation}

with $H_{k}^{II}=\left[H_{k}^{I}\right]^{*}$. The basis is $\psi_{k}^{\dagger}=\left( a_{k}^{\dagger},b_{-k},a_{-k},b_{k}^{\dagger}\right)$. The first block of the Hamiltonian matrix is:

\begin{align}
H^{I}_k&=\left(\begin{array}{cc}
J_{1}S\left(1-\gamma_{k}\right)+J_{2} s+A\tilde{S} & \sqrt{sS} \left(  J_{2}\eta_{k}-2iDm_{k} \right)\\
\sqrt{sS} \left( J_{2}\eta_{k}+2iDm_{k} \right) & J_{2} S
\end{array}\right)  \nonumber \\
 &\equiv \left(\begin{array}{cc}
r\left(\mathbf{k}\right) + \Delta\left(\mathbf{k}\right) & f^*\left( \mathbf{k} \right)\\
f\left( \mathbf{k} \right) & r\left(\mathbf{k}\right) - \Delta\left(\mathbf{k}\right) 
\end{array}\right) 
\end{align}

with $\tilde{S}\equiv(2S-1)/2=1/2$ and structure factors $\gamma_k = cos(k_x/2)$, $\eta_k =cos(k_y/2) $, and $m_k=-sin(k_x/2)sin(k_y/2)$.

The Hilbert space was doubled in this procedure, so it carries not only the magnon (pseudo-) particle states but also non-physical hole states. To diagonalize Hamiltonian (\ref{hamiltonian_main}), we use a Bogoliubov transformation, where the two physical solutions have eigenvalues:

\begin{align}
\omega_{\uparrow \downarrow}\left(\mathbf{k}\right)=\sqrt{r\left( \mathbf{k} \right)^{2}- \left| f\left( \mathbf{k} \right) \right|^2}\mp\Delta\left( \mathbf{k} \right)
\end{align}

That is the system's band structure, showing the energies of the bands. As we made $\hbar=1$, the energies in all band structure plots here are represented in units of frequency (hertz). The up/down magnons carry magnetic dipole momentum $\sigma g_{\sigma} \mu_B$, with $\sigma=\pm1$. For AFM systems, the g-factor $g_{\sigma}$ is the same for both magnon species, so it is usually absorbed into another constant. For ferrimagnets, however, usually $g_\uparrow \neq g_\downarrow$ due to the inequivalence of the spins in the two sublattices, and it is essential to carry the g-factors in the expressions \cite{Nakata2021}.

We must remember that magnons are bosons, so there is no Fermi level. The thermal population $n_{\lambda} \left( \mathbf{k} \right)$ is given by the Bose-Einstein distribution: $n_{\lambda}\left( \mathbf{k} \right)=\left( e^{\hbar \omega_{\lambda} \left( \mathbf{k} \right) /k_B T}-1 \right)^{-1}$. Both bands are always populated. Even if a gap exists between the bands, the system is not a true insulator (this term can be applied only as an analogy).

\begin{figure}[h!]
\centering
\includegraphics[width=0.45\textwidth]{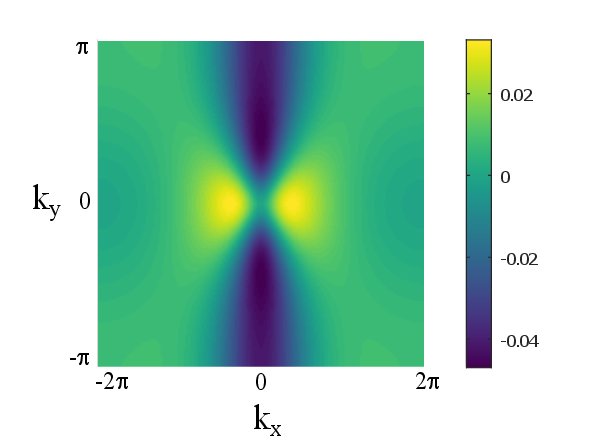} \caption{Berry curvature of both bands throughout the Brillouin zone. Same parameters as Fig. \ref{figure_bands}.} \label{figure_berry}
\end{figure}

\section{Berry curvature and transverse transport}
\label{section_berry}

\begin{figure}[h!]
\centering
\includegraphics[width=0.4\textwidth]{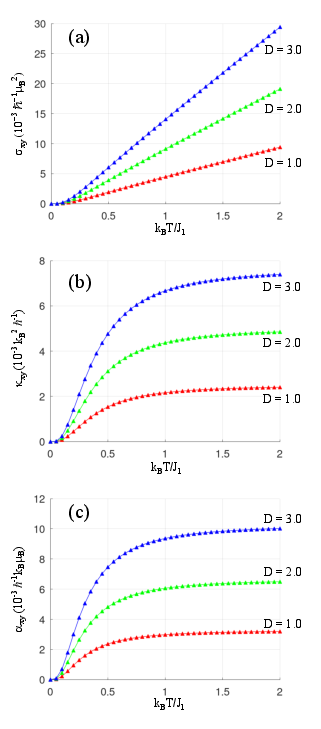} \caption{(a) Spin Hall conductivity $\sigma_{xy}$, (b) thermal Hall conductivity and (c) spin Nernst coefficient as functions of $k_B T/J_1$. The parameters are $J_{1}=J_{2}=1.0$, $A=0.1$, $g_\downarrow=1.0$, $g_\uparrow=1.2$ and three values of $D$.} \label{coeff_D}
\end{figure}

The Berry curvature of the $\lambda$-band $\mathbf{\Omega_{\lambda}}\left(  \mathbf{k}\right)$ is a property of the energy band responsible (among other things) for transversal transport and topological effects. It can be obtained from the eigenstates $u_{\lambda}\left(  \mathbf{k}\right)$ of the Hamiltonian \cite{Teixeira_Pires2019-tx}:

\begin{equation}
\mathbf{\Omega_{\lambda}}\left(  \mathbf{k}\right)  =i\left\langle \nabla_{\mathbf{k}%
}u_{\lambda}\left(  \mathbf{k}\right)  \right\vert \times\left\vert
\nabla_{\mathbf{k}}u_{\lambda}\left(  \mathbf{k}\right)  \right\rangle
\end{equation}

The Berry curvature acts like a fictitious magnetic field on the magnons and is related to the geometrical Berry phase
accumulated by the ground state eigenfunctions when evolving in the
Brillouin zone. In the case of a two-band AFM/FiM Hamiltonian, both bands have the same Berry curvature, whose off-plane component can be obtained analytically from

\begin{align}
\Omega_{\uparrow \downarrow}(\mathbf{k})&=-\frac{1}{2}\sinh\theta_k\left(
\frac{\partial\phi_k}{\partial k_{x}}\frac{\partial\theta_k}{\partial k_{y}}%
-\frac{\partial\phi_k}{\partial k_{y}}\frac{\partial\theta_k}{\partial k_{x}%
}\right)  \\
\end{align}

where $\theta_k$ and $\phi_k$ are parameters that can be written in terms of $r\left( \mathbf{k} \right)$, $\Delta\left( \mathbf{k} \right)$ and $f\left( \mathbf{k} \right)$ \cite{deOliveira2023_Lieb}. 

When an external in-plane field gradient is applied to some systems, it is possible to observe magnon transport in a direction transverse to the field gradient. This phenomenon can be generically called the \textit{Hall-like transport of magnons}. When the external perturbation is a magnetic field gradient we observe a transverse spin current given by \cite{Fujimoto2009,Han2017}:

\begin{equation}
j_y^{S,B}= \sigma_{xy} \left( - \partial_x B \right)
\end{equation}

where $\sigma_{xy}$ is the spin Hall conductivity, and this phenomenon is called the \textit{spin Hall effect} of magnons.

When we apply a temperature gradient, we observe both spin and thermal transverse currents, given respectively by \cite{Katsura2010,Matsumoto2014,Han2017}:

\begin{align}
j_y^{S,T} &= \alpha_{xy} \left( - \partial_x T \right) \\
j_y^{Q,T} &= \kappa_{xy} \left( - \partial_x T \right)
\end{align}

These are the \textit{spin Nernst effect} and the \textit{thermal Hall effect} of magnons. The coefficient $\alpha_{xy}$ is called the spin Nernst coefficient, and $\kappa_{xy}$ is the thermal Hall conductivity. Within the linear response theory, the transport coefficients for each band are given by integrals (in the continuum limit) that involve the Berry curvature \cite{Nakata2021,Nakata2017}:

\begin{align}
[\sigma_{xy}]_{\uparrow \downarrow}&=-\frac{\left(g_{\uparrow \downarrow}\mu_B\right)^2}{\hbar V_{BZ}} \int\limits_{BZ} d^2k \,\,\, n_{\uparrow \downarrow}\left( \mathbf{k} \right)\Omega_{\uparrow \downarrow}(\mathbf{k}) \label{cond}\\
[\alpha_{xy}]_{\uparrow \downarrow}  &  =-\frac{\left(g_{\uparrow \downarrow}\mu_B\right)k_{B}}{\hbar V_{BZ}} \int\limits_{BZ} d^2k \,\,\, c_1 \left[ n_{\uparrow \downarrow} \left( \mathbf{k} \right) \right] \Omega_{\uparrow \downarrow}(\mathbf{k}) \label{alpha}\\
[\kappa_{xy}]_{\uparrow \downarrow}  &  =-\frac{k_{B}^{2}T}{\hbar V_{BZ}} \int\limits_{BZ} d^2k \,\,\, c_2 \left[ n_{\uparrow \downarrow} \left( \mathbf{k} \right) \right] \Omega_{\uparrow \downarrow}(\mathbf{k})  \label{kappa}%
\end{align}

Here, $n  \left( \mathbf{k} \right) $ is the thermal population of the band given by the Bose-Einstein distribution, and the functions $c_1$ and $c_2$ are defined as

\begin{align}
c_{1}(x)  &=\left(  1+x\right)  \ln\left(  1+x\right)
-x\ln\left(  x\right) \\
c_{2}(x) &=\left(  1+x\right)  \left[  \ln\left(
\frac{1+x}{x}\right)  \right]  ^{2}-\left(  \ln x\right)  ^{2}-2Li_{2}\left(-x\right) 
\end{align}

where $Li_{2}\left(  x\right)$ is Spence's dilogarithm function.

It is important to remark that in Refs. \cite{Nakata2017} and \cite{Nakata2021} the authors write the transport coefficients without the explicit integrals and in terms of the Chern number of the band. In those references it is assumed that the bands are almost flat and the functions of $n  \left( \mathbf{k} \right)$ are factored out of the integrals in Eqs. (\ref{cond})-(\ref{kappa}), so the coefficients become proportional to the Chern number $C=\int d^2k \,\, \Omega/2\pi$. We do not make the flat band approximation here, so the functions of $n  \left( \mathbf{k} \right)$ are not factored out.

The transverse currents of both magnons combine to generate the total conductivities of the system:

\begin{align}
\sigma_{xy}&=[\sigma_{xy}]_\downarrow+[\sigma_{xy}]_\uparrow \label{cond_total} \\
\alpha_{xy}&=[\alpha_{xy}]_\downarrow-[\alpha_{xy}]_\uparrow \label{alpha_total}\\
\kappa_{xy}&=[\kappa_{xy}]_\downarrow+[\kappa_{xy}]_\uparrow \label{kappa_total}
\end{align}

The difference in sign between $\alpha_{xy}$ and $\kappa_{xy}$ can be understood as follows. The thermal current is additive and the spin current is subtractive if a perturbation drives both magnons in the same direction, and vice-versa if it drives the magnons in opposite directions (which is the case of a thermal gradient in the system studied here). Furthermore, in degenerate systems both currents have the same magnitude, and we can observe a pure spin current without thermal current (\textit{pure spin Nernst effect of magnons}) \cite{Cheng2016,Zyuzin2016,Nakata2017,deOliveira2023_2}. These systems are $\mathbb{Z}_2$ topological magnon insulators, protected by an effective time-reversal symmetry. They are analogs of the quantum spin Hall states of electrons \cite{KaneMele}.

\begin{figure}[ht!]
\centering
\includegraphics[width=0.45\textwidth]{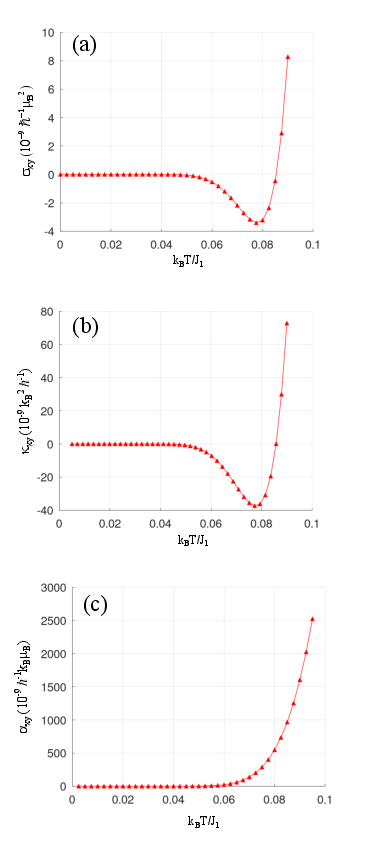} \caption{(a) Spin Hall conductivity $\sigma_{xy}$, (b) thermal Hall conductivity and (c) spin Nernst coefficient as functions of $k_B T/J_1$. The parameters are $J_{1}=J_{2}=1.0$, $D=0.3$, $A=1.1$, $g_\downarrow=1.0$, and $g_\uparrow=1.2$. Note that the first two coefficients changes for low temperatures, which is a result of the competing conductivities of each individual band.} \label{coeff_detalhe}
\end{figure}

\begin{figure}[h!]
\centering
\includegraphics[width=0.45\textwidth]{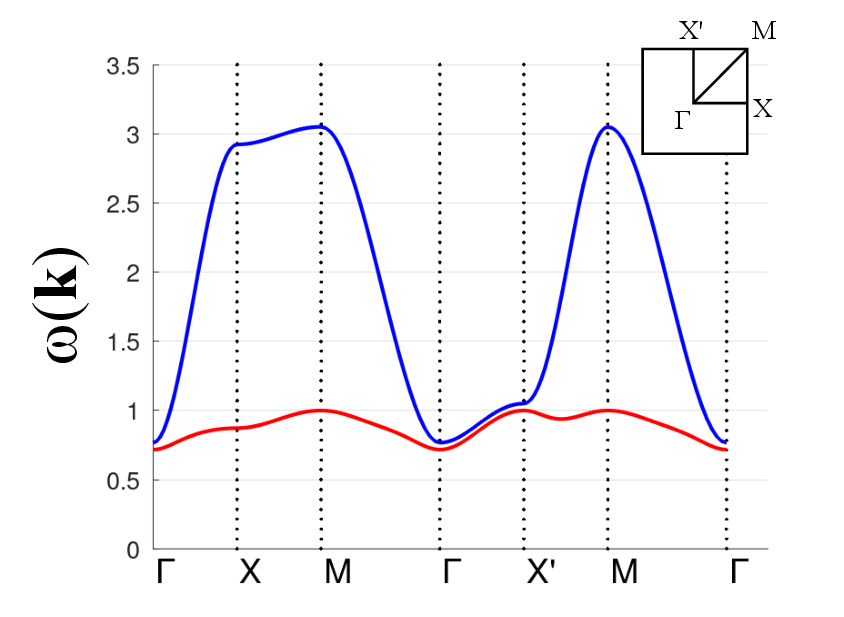} \caption{Band structure for $J_{1}=J_{2}=1.0$, $D=0.3$, and $A=1.1$. For $A>J_2$ the system is gapped.} \label{figure_bands_detalhe}
\end{figure}

\section{Results}
\label{section_results}

The band structure of the system can be seen in Figure \ref{figure_bands}. For the typical values of $A<J_2$ the bands cross in two separate paths in the Brillouin zone. The Berry curvature of the system is represented in Figure \ref{figure_berry}. The symmetrical shape of the Berry curvature shows us that its integration in the Brillouin zone is zero, so the system is not a Chern insulator ($C=0$). We can see that the points of higher (positive) Berry curvature are located in the $\Gamma-X$ path, in regions where $\omega_\downarrow$ dominates. On the other hand, the points of lower (negative) Berry curvature lay in the $\Gamma-X'$ path, corresponding to points where $\omega_\uparrow$ dominates. The result is that, from Eqs. (\ref{cond})-(\ref{kappa}), the transport coefficients are positive for the $\downarrow$-band and negative for the $\uparrow$-band. We see that the total spin Hall conductivity and thermal Hall conductivity (Eqs. (\ref{cond_total}) and (\ref{kappa_total})) become subtractive, while the spin Nernst coefficient (Eq. (\ref{alpha_total})) becomes additive. That agrees with the literature, which states that a magnetic field gradient drives both species of magnons in the same transverse direction, so the spin currents compete with each other. Also, a temperature gradient drives the magnons in opposite directions, so the spin currents are reinforced, and the thermal current is subtractive \cite{Nakata2017}. For some collinear antiferromagnetic systems ($\mathbb{Z}_2$ topological magnon insulators), an effective time-reversal symmetry results in degenerate bands, and we can have a transverse spin current without heat current, realizing a magnonic version of the quantum Hall effect \cite{Cheng2016,Zyuzin2016,Nakata2017,deOliveira2023_2}. Such symmetry does not exists in a ferrimagnet, and the degeneracy is broken \cite{Park2020,Nakata2021,Ohnuma2013}. We have non-null and tunable heat and spin currents.

The transport coefficients versus temperature are plotted in Figure \ref{coeff_D}. We see the well-known asymptotical behavior of $\alpha_{xy}(T)$ and $\kappa_{xy}(T)$. The coefficients increase with $D$, and vanish when $D=0$ (as well as the Berry curvature), meaning that the Dzyaloshinkii-Moriya interaction is responsible for the transverse transport. All three coefficients are non-null, as expected from a non-degenerate system.

For $A \geq J_2$ it is possible to observe an interesting behavior in the $\sigma_{xy}$ and $\kappa_{xy}$ (subtractive) coefficients: the curves are non-monotonic and present a sign change, while that does not occur $\alpha_{xy}$ (Fig. \ref{coeff_detalhe}). That can be explained as follows. When $A \geq J_2$, the narrower $\uparrow$-band doesn't cross the wider $\downarrow$-band (Figure \ref{figure_bands_detalhe}). We know that in low temperatures the lower band dominates, so in this case, the narrower band (which has negative $\left[\sigma_{xy}\right]_\uparrow$ and $\left[\kappa_{xy}\right]_\uparrow$) is more populated, resulting in negative total transport coefficients. As the temperature rises, the population in the wider band surpasses the narrower one, and the total $\sigma_{xy}$ and $\kappa_{xy}$ become positive. This competition does not occur for $\alpha_{xy}$, which is additive. That behavior of $\sigma_{xy}$ and $\kappa_{xy}$ does not occur for $A<J_2$, as the wider band reaches lower values, and its population dominates in all temperatures. The non-monotonic character of the curves does not change with the inclusion of NN exchange between the $s=1/2$ sites. However, the valley becomes rapidly less pronounced and closer to $T=0$ with the rise of the exchange parameter. We also investigated the role of different spins in the lattice, and the conclusion was that the curves can be non-monotonic independently of spin value for both sites, provided that at least one spin is different from $1/2$. That occurs even for $S=s$ (the AFM case), when the bands touch but do not cross, for all values of $A$.

In summary, the necessary conditions for the non-monotonic behavior are: (1) The SIA exists and is effective ($A \neq 0$ and at least one spin is different from $1/2$), and (2) the bands do not cross. This behavior seems to be an intrinsic feature of the square geometry in the C-type ordering investigated here. The change of sign that follows the non-monotonic transport coefficients, which also happens in other magnon systems, opens the exciting possibility of controlling the direction of transverse magnon flow with the temperature.

\section{Conclusion}
\label{section_conclusion}

We have studied a 2D magnetic lattice that describes a layer of the $Cu_2F_5$ crystal \cite{Rybin2021,Korotin2021,Korotin2023} using the spin wave approach. The magnetic order is ferrimagnetic, with spin-up sites ($S=1$) pointing in the off-plane direction opposite to spin-down ($s=1/2$). The magnon bands are not degenerate, as expected for ferrimagnets, and in contrast to AFM collinear order in the Néel state, where an effective time-reversal symmetry makes the bands degenerate. A non-null Berry curvature is generated by the Dzyaloshinskii-Moriya interaction between next-near neighbors. That enables the system to show transverse transport effects. An interesting remark is that, for high single-ion anisotropy ($A\ \geq J_2$), the directions of net spin (in response to a magnetic field gradient) and heat (in response to a temperature gradient) currents are sensitive to temperature,resulting in non-monotonic curves for spin Hall coefficient and thermal Hall coefficient. That can be technologically interesting regarding the control of magnon flow. The non-monotonic curves seem to be an intrinsic feature of the system regardless of spin value, happening even when both spins are equal, provided that a single-ion anisotropy is present (and effective) and the magnon bands do not cross.

\section{Acknowledgments}

This work was supported by CAPES (Coordenação de Aperfeiçoamento de Pessoal de Nível Superior) and CNPq (Conselho Nacional de Desenvolvimento Científico e Tecnológico).

\bibliographystyle{elsarticle-num} 
    \footnotesize   
   	\bibliography{refs}

\end{document}